\documentstyle[aps]{revtex}
\input epsf

\def\f{\bbox{f}}

\begin{document}

\draft
\title{Observing Long Cosmic Strings Through Gravitational Lensing}
\author{Andrew A. de Laix}
\address{Case Western Reserve University, Department of Physics,
Cleveland Ohio, 44106-7079}
\date{\today}
\maketitle
\begin{abstract}
We consider the gravitational lensing produced by long cosmic strings
formed in a GUT scale phase transition.  We derive a formula for the
deflection of photons which pass near the strings that reduces to an
integral over the light cone projection of the string configuration
plus constant terms which are not important for lensing.  Our strings
are produced by performing numerical simulations of cosmic string
networks in flat, Minkowski space ignoring the effects of cosmological
expansion.  These strings have more small scale structure than those
from an expanding universe simulation---fractal dimension 1.3 for
Minkowski versus 1.1 for expanding---but share the same qualitative
features.  Lensing simulations show that for both point--like and
extended objects, strings produce patterns unlike more traditional
lenses, and, in particluar, the kinks in strings tend to generate
demagnified images which reside close to the string.  Thus lensing
acts as a probe of the small scale structure of a string.  Estimates
of lensing probablity suggest that for string energy densities
consistant with string seeded structure formation, on the order of
tens of string lenses should be observed in the Sloan Digital Sky
Survey quasar catalog.  We propose a search strategy in which string
lenses would be identified in the SDSS quasar survey, and the string
nature of the lens can be confirmed by the observation of nearby high
redshift galaxies which are also be lensed by the string.
\end{abstract}

\section{introduction}
\label{introduction}
The current understanding of the formation of large scale structure
presumes that clusters, walls, filaments, etc.\ result from the
gravitational collapse of small deviations from homogeneity in the
early universe, but the source of these fluctuations is, at present,
still a mystery.  The most popular model is inflation, where vacuum
energy in the early universe drives a period of exponential growth
which stretches small quantum mechanical fluctuations to macroscopic
scales much greater than the horizon at that time.  As the universe
expands, these fluctuations fall back into the causal horizon and
collapse into the structure we see today.  However, there is yet no
definitive evidence in either accelerator experiments or cosmological
observations to prove that a field with the appropriate
characteristics exists to induce such an inflationary epoch.  It is
worthwhile, then, to consider alternatives to inflation which can also
explain the origin of structure, specifically, topological defects.
During cosmological phase transitions, topological defects can form
where adjacent fields take on different vacuum states which can only
be continuously connected if a region of false vacuum is trapped
between them.  The energy trapped in these defects can then
gravitationaly induce perturbations which will collapse into the
observed structure.  A particularly interesting subclass of defect
models is cosmic strings, which are lineal structures formed when a
$U(1)$ symmetry is spontaneously broken.  If these strings were formed
at the GUT scale, they can have sufficient energy, and therfore mass,
to drive the necessary perturbations in the ordinary matter which 
collapse into the structure we observe today.  The question then
becomes, how can we observe strings and compare these models with
inflation?  Inflationary models may best be tested by microwave
background observations, as these models make precise predictions for
the spectrum of microwave perturbations from $180^{\circ}$ scales down
to several orders of magnitude smaller.  For string models, things are
less certain.  On large scales, strings predict a spectrum similar
enough to inflation as to be statistically indistinguishable
\cite{allen}.  On smaller scales, issues like
decoherence\cite{magajio}, which smooth the acoustic peaks, are not
fully understood, and while it is likely that we could rule out
inflation if strings are the true model, we may not be able to confirm
the existence of strings through microwave observations, given the
current uncertianty about the CMBR fluctuations they produce.  In this
paper, we shall discuss another method by which cosmic strings can be
observed, namely through gravitational lensing.  Unlike other defect
models, where we expect perhaps only one defect to remain in a horizon
volume, simulations suggest that a significant length of string is
observable today.  The gravitational fluctuations these strings induce
will bend light, making it possible to observe a cosmic string if it
is back lit by a visible source. Since there are many sources,
including galaxies and quasars, which can light up a string, we will
consider the structure of images that would arise in such lensing
systems containing long cosmic string.  Our estimates of the string
lensing probability suggest that string lenses occur at a rate
of 10\% to 30\% of that of galaxy lenses in the case of quasar
sources, a few dozen of which have been observed.  We show that the
resulting images have a unique signature, that is, along with the
image pair that is associated with infinite straight strings, there
will usually be a series of smaller, demagnified images which reside
closer to the string itself.  We conclude that new quasar surveys with
large angular sky coverage, like the Sloan Digital Sky Survey, should
contain a significant number of string lenses and be able to
definitively observe or rule out the cosmic string model, at least for
flat space.

This paper is organized into several sections.  In the following, we
estimate the lensing probability for long cosmic strings and compare
it to that for galaxies.  In \S \ref{deflection} we calculate the
deflection of a photon in the presence of a long cosmic string.  In \S
\ref{lenstheory}, we discuss the basic theory of gravitational
lenses. The next section is broken into two parts, where \S
\ref{network} discusses the generation of the long strings used in our
calculations and \S \ref{images} explains how these were used to find
lensed images.  Finally, in \S \ref{conclusion} we discuss our results
and suggest a search strategy for finding string lenses.   

\section{Long String Lensing probability}
\label{probability}
We begin by considering the likelihood observing a string lens
system. Quasars represent the best objects for seeing long string
lensing, as they are bright, with images not likely to be lost in any
background, and new surveys like the Sloan Digital Sky Survey will
observe them in large numbers (on the order of $10^5$ quasars over 1/4
of the sky).  We would like to know how many quasar--string lensing
systems we can expect to observe in such a survey, but this depends on
how a particular quasar sample is observed.  A simpler calculation is
to measure the optical depth for lensing, that is the the probability
of an object at a given redshift being lensed, which we can then
compare to optical depths for quasar--galaxy lensing, which has
already been observed for about a dozen systems and should occur on
the order of hundreds of cases in the SDSS.

To estimate the string lensing probability, we require two pieces of
information: the projected angular density of string and a lensing
cross section for that string.  With regards to the former, numerical
simulations of string networks in an expanding universe can give
reliable estimates of the string density $\rho_{ls}$ in a horizon
volume, characterized by the horizon radius $d_H$.  Using the results
of Bennett and Bouchet\cite{bennett} as a representative example, we
see that in the matter dominated epoch the length of string in horizon
units is a constant given by
%
%
\begin{equation}
\label{lstring}
L_{ls} = {\rho_{ls}d_H^2 \over \mu} = 31 \pm 7,
\end{equation}
where $\mu$ is the energy per unit length of string.  We shall treat
this string as an ensemble of small links, with length $L_l$ in
horizon units, distributed with a uniform probability density related
to the above result, each with a random orientation with respect to
the line of sight.  These assumptions are appropriate when considering
an average over many string networks.  We shall also assume that each
of the links is static, since we lack good information on the
distribution of link velocities. In essence we are disregarding the
effects of the Lorentz contraction (see next section for details),
which means our estimate is probably only accurate to a factor of a
few.  We now subdivide space into differential volume elements such
that the probability that more than one link, located by its center of
mass, resides in the same volume is vanishingly small in comparision
to the probability that one link resides in that volume.  For the
$i$th volume element, we find that the differential angle of sky
subtended by the string is given by
%
%
\begin{equation}
\label{dangle}
d\Theta_i = n_i {L_l d_H\over d_A} \sin(\beta_i) dV_i,
\end{equation}
where $n_i$ is the number of links in the volume, $\beta_i$ is a random
orientation angle associated with each link, and $d_A$ is the angular
diameter distance to the link
%
%
\begin{equation}
\label{angdia}
d_A = {2 \over H_0 (1+z)^2} (1+z-\sqrt{1+z}),
\end{equation}
assuming a flat, matter dominated universe.  Under the same
assumptions, the volume element and horizon distance can also be
expressed in terms of the redshift:
%
%
\begin{eqnarray}
dV &=& {4 \over H_0^3} {(1+z - \sqrt{1+z})^2 \over (1+z)^3 \sqrt{1+z}}
dz d\Omega, \\ \nonumber
d_H &=& {2 \over H_0 (1+z)^{3/2}}.
\end{eqnarray}

To estimate the lensing cross section, we assume that each link has
the same cross section as if it were part of an infinite straight
string.  In this case, the angular cross section per radian of string
is
%
%
\begin{equation}
\label{cross}
\delta \phi = 8\pi G \mu {D_{lq} \over D_q} \sin{\beta}.
\end{equation}
The distances $D_{lq}$ and $D_q$ are respectively the angular diameter
distances from the lens to the quasar and from the source to the
quasar.  In flat space for a source located at a redshift of $z_2$ and
an observer located at $z_1$, the general form of the angular diameter
distance is given by
%
%
\begin{equation}
\label{angulardiameter}
d_A(z_1,z_2)= {2 \over H_0 (1+z_2)} \left[ {1 \over \sqrt{1+z_1}}-{1
\over \sqrt{1+z_2}} \right]. 
\end{equation}
We ignore the effects of structure formation on theses distances which
introduce deviations from homogeneity that can effect the path length.
Convolving the cross section with the angular string density, we can
determine the optical depth for a quasar at a redshift of $z = z_q$:
%
%
\begin{equation}
\label{optical}
\tau(z_q) = 8\pi G\mu L_l{1\over \Omega_{O}} \sum_i n_i \sin^2(\beta_i)
{d_H \over d_A} {D_{lq} \over D_q},
\end{equation}
where $\Omega_{O}$ is the observed fraction of sky.  Taking the
expectation value for $\tau$, we find that $\langle n_i\rangle = L_{ls}/(L_l
d_H^3)$ and $\langle\sin^2(\beta)\rangle = 2/3$.  This leaves us with the
integral over the observed volume
%
%
\begin{equation}
\label{optical1}
\langle \tau \rangle  =  {16 \over 3} \pi G\mu L_{ls} \int {dV \over
\Omega_0} {1 \over d_A d_H^2} {D_{lq} \over D_q},
\end{equation}
which can be performed analytically, giving the final result:
%
\begin{equation}
\label{depth}
\tau(z_q) = {8 \over 3} \pi G \mu L_{ls} \left[{1 \over 21} z_q^3
+ {13 \over 105} z_q^2 + {3 \over 35}
z_q + {2 \over 35} - \sqrt{1+z_q}\left( {2 \over 105} z_q^2 +{2 \over
35} z_q +{2 \over 35} \right)\right].
\end{equation}

Now we would like to estimate the variance in $\tau$ which will permit
us to set limits on string parameters using lensing statistics.
Results from numerical simulations indicate that on scales of about
$0.1 d_H$, the string undergoes a random walk.  To make the problem
tractible, we shall continue to treat each of these segments as
uncorrelated with the rest, and presume that the random walk
correlation is fairly small.  The mean of the square of the
optical depth is given by
%
%
\begin{equation}
\label{tausq}
\left \langle \tau^2 \right\rangle = {1 \over \Omega_O^2}(8\pi G\mu L_l)^2
\sum_{i,j} \langle n_i n_j \rangle  \langle \sin^2(\beta_i)
\sin^2(\beta_j)\rangle
 \left({d_h
\over d_A}{D_{lq} \over D_q}\right)_i \left({d_h\over d_A}{D_{lq}
\over D_q}\right)_j.
\end{equation}
The expectation value $\langle n_i n_j \rangle$ in the limit of differential
volumes is equal to 
%
%
\begin{equation}
\label{ninj}
\langle n_i n_j \rangle = n^2 dV_i dV_j + n dV_i \delta_{i,j},
\end{equation}
when $n_i$ and $n_j$ are uncorrelated, and $n = \langle n_i \rangle$.
Given this relation, one can easily show that
%
%
\begin{equation}
\label{variance}
\sigma_\tau^2 = 
\langle \tau^2\rangle - \langle\tau\rangle^2 = 
{1 \over \Omega_O^2}(8\pi G\mu L_{ls})^2 {L_l
\over L_{ls}} \int dV {8 \over 15} {1 \over d_H d_A^2}
\left({D_{lq} \over D_q}\right)^2,
\end{equation}
which has the analytic result
%
%
\begin{equation}
\label{varianceAnalytic}
 \sigma_\tau^2  = {1 \over \Omega_O}(8\pi G\mu L_{ls})^2 
{L_l\over L_{ls}} {2 \over 225} { z_q^3+3z_q^2-12z_q-24+24\sqrt{1+zq}
\over (\sqrt{1+z_q}-1)^2}.
\end{equation}
To get the full error in $\tau$ we convolve this result with the
theoretical uncertianty in the long string density $\sigma_{ls}$ given
in eq.\ (\ref{lstring}), which adds in quadrature.  Now consider a
distribution of sources with mean number density as a function of
redshift $N(z)$.  The expected number of observed lenses is given
simply by
%
%
\begin{equation}
\label{meanLens}
\int dV N(z) \tau(z). 
\end{equation}
To calculate the variance, we must account for both the variance in
$\tau$ and the Poisson fluctuations in the source distribution.
Including both these effects, we find the full variance is
%
%
\begin{equation}
\label{varianceLens}
\sigma^2 = \int dV \left( N(z)\tau(z) + N(z)^2\sigma_\tau^2(z) +
N(z)^2 \tau^2(z) {\sigma_{ls}^2 \over L_{ls}} \right). 
\end{equation}
As a toy example, we consider a quasar distribution given by $N(z) =
\delta(z-2) 10^5/\pi$, which rough approximates that of the Sloan
Digital Sky Survey.  The SDSS will observe one quarter of the full
sky, so using the results of this section, we find for $G\mu = 
10^{-6}$, that the number of observed quasar lenses is $18 \pm
6$. 
  
\begin{figure}
\caption{\protect\label{tau} The probability of a source being lensed
by a long cosmic string plotted as a function of source redshift
(solid).  For comparison, the probability of a source being lensed by
a galaxy is shown (dashed).
}
\epsfxsize = \hsize \epsfbox{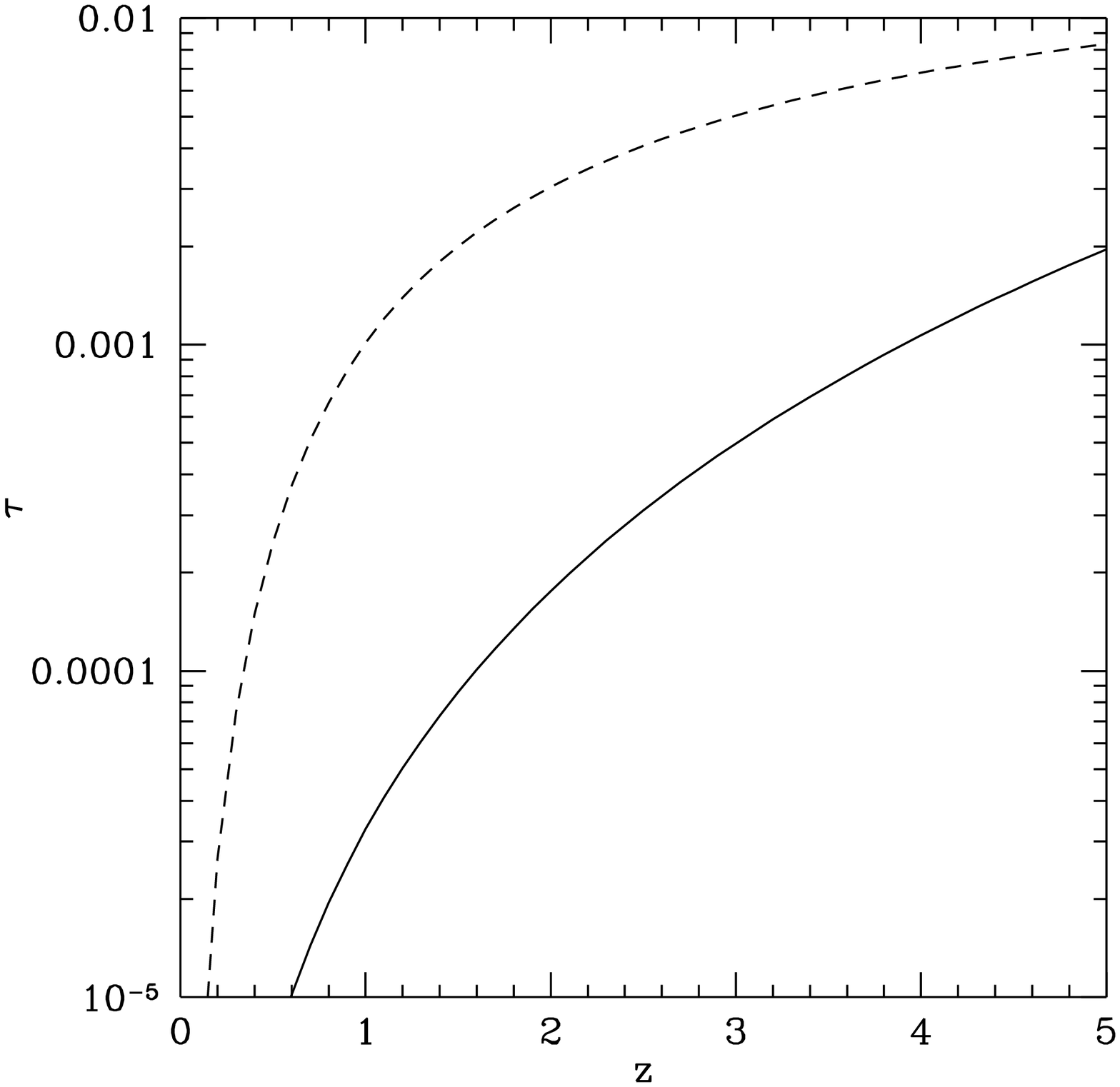}
\end{figure}

In figure \ref{tau} we show the optical depth for long cosmic strings,
assuming a value $G \mu = 10^{-6}$, compared with the optical depth
resulting from galaxy lenses as calculated by Turner, Ostriker, and
Gott \cite{turner}.  The galaxy estimate is based on an isothermal
sphere model, which represent an upper limit to the lensing
probability, as including a finite core to the galaxy tends to reduce
the optical depth by up to 50\% \cite{white}.  Thus we  expect
that for a wide field survey anywhere from 10\% to 30\% of the
observed lenses will be the result of long cosmic strings when
compared to galaxy lenses (assuming $G\mu = 10^{-6}$).  Given that
order 10 quasar lensed have been observed, one may expect that a few
string lenses should have been seen.  In fact there are no lenses
that have yet been ascribed to cosmic strings, but this is
statistically unsuprising because of the large variance associated
with small number statistics.  We can, however, reasonably conclude
that string tensions $G\mu$ greater that a few $10^{-6}$, which would
predict tens of observed lenses, probably are ruled out, consistant
with the limits from pulsar timing.

\section{Geodesic Deflection By A Long Cosmic String}
\label{deflection}
In a previous paper\cite{delaix}, we derived an equation for the
deflection of a null geodesic--corresponding to a photon trajectory in
the geometric limit--arising from the gravitational field of a cosmic
string loop.  For long cosmic strings, we can make use of some of
that derivation, but we must now be careful, as the long strings
stretch to horizon scales, requiring us to consider certain surface
terms which could safely be ignored when examining small loops.  Let
us begin again by assuming a weak field so that the full metric may be
expressed as
%
%
\begin{equation}
\label{metric}
g_{\mu \nu} = \eta_{\mu \nu} + h_{\mu \nu},
\end{equation}
where $\eta_{\mu \nu} = {\rm diag}(-1,1,1,1)$ is the usual Minkowski
metric and $ h_{\mu \nu}$ is a small perturbation such that all terms
of $O(h^2)$ are negligible.  For simplicity, we choose to work in the
harmonic gauge, which implies the condition $g_{\mu \nu}
\Gamma^\lambda_{\mu \nu} = 0$.  Using this gauge choice, to linear
order we are left with a simple wave equation for the metric
%
%
\begin{equation}
\label{wave}
\Box^2 h_{\mu \nu} = -16\pi G  S_{\mu \nu},
\end{equation}
where $S_{\mu \nu} = T_{\mu \nu} - 1/2 \eta_{\mu \nu}
T^\lambda_\lambda$ is the traceless component of the stress energy
tensor $T_{\mu \nu}$.  If we decompose the photon four velocity
$\gamma^\mu$ into its zeroth and first order pieces, $\gamma^\mu_{(0)}$
and $\gamma^\mu_{(1)}$ respectively, then it is a straightforward
calculation to solve the geodesic equation and show that the
first order deflection for a photon emitted at $t_1$ and observed at
$t_2$ is given by 
%
%
\begin{equation}
\label{geodesicsoln}
\gamma_{\alpha (1)} = {1 \over 2} \int_{t_1}^{t_2} dt~h_{\mu \nu,
\alpha}\gamma_{(0)}^\mu\gamma_{(0)}^\nu - \left.h_{\mu
\alpha}\gamma_{(0)}^\mu\right|_{t_1}^{t_2},
\end{equation}
Where we are integrating over the photons zeroth order trajectory,
{\it i.e.}~$x_\mu = x_{\mu0}+\gamma_{\mu(0)} t$.  For loops, which
are compact, we could safely ignore the second surface term, but for
now we should retain it when considering long cosmic strings as they
are not compact objects.  

To make further progress, we need to consider the form of the stress
energy tensor resulting from a cosmic string.  Strings are well
approximated as lineal gravitational sources, so the string
configuration at any time is given by a two parameter vector function
$\bbox{f}(\sigma,t)$ where $t$ is the time and $\sigma$ is a parameter
which runs along the conformal length of the string.  One may straight
forwardly infer from this that the string traceless stress energy may
be written in the form
%
%
\begin{equation}
\label{stress}
S_{\mu \nu} = \mu \int d\sigma~F_{\mu \nu}
\delta^{(3)}(\bbox{x}-\bbox{f}(\sigma, t)),
\end{equation}
Where $\mu$ is again the string energy density and $F_{\mu \nu}$ can
be expressed in terms of $\bbox{f}$ and its derivatives.  Now, let us
designate $\tilde{\gamma}_\alpha$ to be the contribution to the first
order deflection which comes from the integral in eq.\
(\ref{geodesicsoln}), separating it from the surface term.
Contracting this with a derivative, we get
%
%
\begin{eqnarray}
\label{gammaderiv}
\partial^{\alpha} \tilde{\gamma}_\alpha &=& {1 \over 2}
\int_{t_1}^{t_2} dt~\Box^2 h_{\mu \nu}
\gamma_{(0)}^\mu\gamma_{(0)}^\nu  \nonumber \\
&=& -8 \pi G \int_{t_1}^{t_2} dt~S_{\mu \nu}\gamma_{(0)}^\mu\gamma_{(0)}^\nu,
\end{eqnarray}
where the second line comes from the metric equation (\ref{wave}).
Plugging in the stress energy from eq.\ (\ref{stress}), we see that
%
%
\begin{equation}
\label{gammaderiv1}
\partial^{\alpha} \tilde{\gamma}_\alpha =  -8 \pi G \mu \int d\sigma
\int_{t_1}^{t_2}
dt~F_{\mu \nu}\delta^{(3)}(\bbox{x}_0+\bbox{\gamma}_{(0)}t-
\bbox{f}(\sigma, t)) \gamma_{(0)}^\mu\gamma_{(0)}^\nu.
\end{equation}
We can evaluate the time integral if we decompose $\bbox{f}$ into
components which are perpendicular, $\bbox{f}_\bot$, and parallel,
$\bbox{f_\|}$, to the zeroth order photon trajectory, with the result
%
%
\begin{equation}
\label{gammaderiv2}
\partial^{\alpha} \tilde{\gamma}_\alpha =  -8 \pi G \mu\int
d\sigma\left[ {F_{\mu \nu} \gamma_{(0)}^\mu\gamma_{(0)}^\nu
\over 1 - \dot{f}_\|}
\delta^{(2)}(\bbox{x}_{\bot 0}-\bbox{f}_\bot)\right]_{t=t_0},
\end{equation}
where $t_0$ is the solution to the equation $t_0 =
f_\|(\sigma,t_0)-x_{\| 0}$, {\it i.e.}~the light cone time slice, and
dots refer to derivatives with respect to time. Note that the limits
on the $\sigma$ integral are constrained to the regions where a
solution with $t_1 < t_0 < t_2$ exists. Now we shall write out the
left hand side of the equality in terms of a parallel and
perpendicular gradients, $\partial^\alpha\tilde{\gamma}_{\alpha(1)} =
\nabla_\bot \cdot
\tilde{\bbox{\gamma}}_{\bot(1) } +  \nabla_\| \cdot
\tilde{\bbox{\gamma}}_{\|(1) }$, where the parallel gradient can be written
as $ \nabla_\| \cdot
\tilde{\bbox{\gamma}}_{\|(1) } = \gamma_{\delta (0)}
\gamma_{(0)}^\beta \partial^\delta\tilde{\gamma}_{\beta (1)}$. 
Now we consider the contraction
$\gamma_{(0)}^\beta\tilde{\gamma}_{\beta (1)}$, which from eq.\
(\ref{geodesicsoln}) is
\begin{eqnarray}
\label{dgammadt}
\gamma_{(0)}^\beta
\gamma_{\beta (1)} &=& {1 \over 2} \int_{t_1}^{t_2} dt~\gamma_{(0)}^\beta
h_{\mu \nu,
\beta}\gamma_{(0)}^\mu\gamma_{(0)}^\nu \\ \nonumber
	&=&  {1 \over 2} \int_{t_1}^{t_2} dt~{d \over dt}
h_{\mu \nu}\gamma_{(0)}^\mu\gamma_{(0)}^\nu \\ \nonumber
	&=& \left.{1 \over 2}h_{\mu \nu}\gamma_{(0)}^\mu\gamma_{(0)}^\nu
	\right|_{t_1}^{t_2}.
\end{eqnarray}
We are able to perform this integration because $\gamma_{(0)}^\beta
\partial_\beta$ is equivalant to taking a complete derivative with
time, $d/dt$. Using the above result in conjunction with eq.\
(\ref{gammaderiv2}), it is easy to show 
%
%
\begin{equation}
\label{perpderiv}
\nabla_\bot \cdot \tilde{\bbox{\gamma}}_{\bot(1)} =  - 8 \pi G \mu\int
d\sigma\left[ {F_{\mu \nu} \gamma_{(0)}^\mu\gamma_{(0)}^\nu
\over 1 - \dot{f}_\|}
\delta^{(2)}(\bbox{x}_{\bot 0}-\bbox{f}_\bot)\right]_{t=t_0} 
- \left.{1 \over 2}
{d \over dt} h_{\mu \nu}\gamma_{(0)}^\mu\gamma_{(0)}^\nu\right|_{t_1}^{t_2}.
\end{equation}
We can solve this equation by assuming that the first order deflection
can be written as a gradient of a potential,
$\tilde{\bbox{\gamma}}_{\bot(1)} = \nabla_\bot \Phi$, leaving a two
dimensional Poisson equation from which $\Phi$ may be found by
integrating over the Greens function for the two dimensional
Laplacian, $G(\bbox{x}_\bot,\bbox{x}'_\bot) = -
\ln(|\bbox{x}'_{\bot 0}-\bbox{x}_{\bot 0}|^2)/4\pi$. Specifically, we get
%
%
\begin{eqnarray}
\label{phi}
\Phi &=&   \left.{1 \over 8\pi}\int
d^2x'_\bot \ln(|\bbox{x}_{\bot 0}-\bbox{x}'_\bot|^2){d \over dt} h_{\mu
\nu}(\bbox{x}'_\bot+\bbox{x}_{\|0}+\bbox{\gamma}t,t)\gamma^ \mu_{(0)}
\gamma^\nu_{(0)} \right|_{t_1}^{t_2} \\ \nonumber
&& -2 G \mu \int d\sigma~\left[ {F_{\mu \nu} \gamma^\mu_{(0)}
\gamma^\nu_{(0)}  \over 1-\dot{f}_\|}
\ln(|\bbox{f}_{\bot}-\bbox{x}_{\bot 0}|^2)\right]_{t = t_0}.
\end{eqnarray}
Finally, to recover the perpendicular deflection, we take the gradient
and add the surface term from eq.\ (\ref{geodesicsoln})
which leaves us with
%
%
\begin{eqnarray}
\label{deflecttilde}
\bbox{\gamma}_\bot &=&  {1 \over 4\pi}\left.\int
d^2x'_\bot { \bbox{x}_{\bot 0}-\bbox{x}'_\bot \over |\bbox{x}_{\bot
0}-\bbox{x}'_\bot|^2} {d \over dt} h_{\mu
\nu}(\bbox{x}'_\bot+\bbox{x}_{\|0}+\bbox{\gamma}t,t)
\gamma^\mu_{(0)}\gamma^\nu_{(0)} \right|_{t_1}^{t_2} \\ \nonumber
&& + 4 G \mu \int d\sigma~\left[ {F_{\mu \nu} \gamma^\mu_{(0)}
\gamma^\nu_{(0)}  \over 1-\dot{f}_\|}
{\bbox{f}_{\bot}-\bbox{x}_{\bot 0} \over
|\bbox{f}_{\bot}-\bbox{x}_{\bot 0}|^2}\right]_{t = t_0} -
\left. \bbox{h}_\bot \right|_{t_1}^{t_2}.
\end{eqnarray}
where $\bbox{h}_\bot$ is defined to be the perpendicular part of
$h_{\mu \alpha} \gamma^\mu_{(0)}$.  This result gives us exactly what
we want for lensing calculations, the photon deflection away from its
zeroth order path.  The values of $\gamma_{0}$ and $\gamma_{\|}$ which
are equivalent to first order, give us the redshift of a photon as it
passes a string, but we are not interested in this calculation here.

Eq. (\ref{deflecttilde}) may at first seem to be a retrograde step as
it involves a two dimensional integral of the metric in space where
originally we had only a one dimensional integral of the metric over
time. However, we argue that the first and third terms which contain
the metric explicitly can be neglected. To do so we require an
explicit solution for the metric in terms of the stress energy:
%
%
\begin{equation}
\label{stressmetric}
h_{\mu \nu}(\bbox{x},t) = 4 G \int d^3x' {S_{\mu \nu}(\bbox{x}', \tau)
\over \left| \bbox{x} - \bbox{x}' \right| },
\end{equation}
where $\tau = t - \left| \bbox{x} - \bbox{x}' \right|$ is the retarded
time, and this solution is derived from the Greens function for the
$\Box^2$ operator.  Using our string stress energy given in eq.\
(\ref{stress}), we can reduce the metric to 
%
%
\begin{equation}
\label{stringmetric}
h_{\mu \nu}(\bbox{x},t) = 4 G \mu \int d\sigma {F_{\mu \nu}(\sigma, \tau)
\over \left| \bbox{x} - \bbox{f} \right| - \left( \bbox{x} - \bbox{f}
\right) \cdot \dot{\bbox{f}}}.
\end{equation}
Now consider points which are far from the string.  The metric goes
like an integral over $|\bbox{x} - \bbox{f}|^{-1}$ while the middle
term in eq.\ (\ref{deflecttilde}) goes like an integral over
$|\bbox{x}_{\bot 0} - \bbox{f}_\bot|^{-1}$.  The latter represents the
minimum distance between the zeroth order photon trajectory and the
cosmic string, while the former will, in general, go like the distance
from the photon to the string at the current time.  In the case of
string loops, we could expect that the two metric terms would fall off
like the inverse of the distance while the middle term remained
constant, thus allowing us to drop the explicit metric terms.  With an
infinite string, one must be more careful because the distance from
the string can never be large with respect to the string size.
However, when determining the image structure in lensing, it is not
the absolute deflection of the photons which matters, but rather, it
is the difference in deflection between two nearby rays that counts.
In this case, we see that for photons far from the string, the
difference in the contribution between the two metric terms 
declines with distance while the difference between the static terms
remains constant.  So, for photons which pass nearby the string, that
is a small distance when compared to the source and observer, the
effect of the metric terms then is merely to cause all of the images
to be displaced, but not to alter the shape or relative orientation of
the images.  Thus, for the purposes of determining the structure
of strong lensing due to strings, we may drop the explicit metric
terms all together and write
%
%
\begin{equation}
\label{deflect}
\bbox{\gamma}_\bot = 4 G \mu \int d\sigma~\left[ {F_{\mu \nu} \gamma^\mu_{(0)}
\gamma^\nu_{(0)}  \over 1-\dot{f}_\|}
{\bbox{f}_{\bot}-\bbox{x}_{\bot 0} \over
|\bbox{f}_{\bot}-\bbox{x}_{\bot 0}|^2}\right]_{t = t_0},
\end{equation}
where the omitted terms only add a constant to this result when the
photon impact parameter is small compared to the distances of the
source and observer to the lens.

\section{Basic Gravitational Lensing Theory}
\label{lenstheory}
\begin{figure}
\caption{\protect\label{lensdiag} A schematic representation of a
gravitational lensing system.
}
\epsfxsize = 6 in \epsfbox{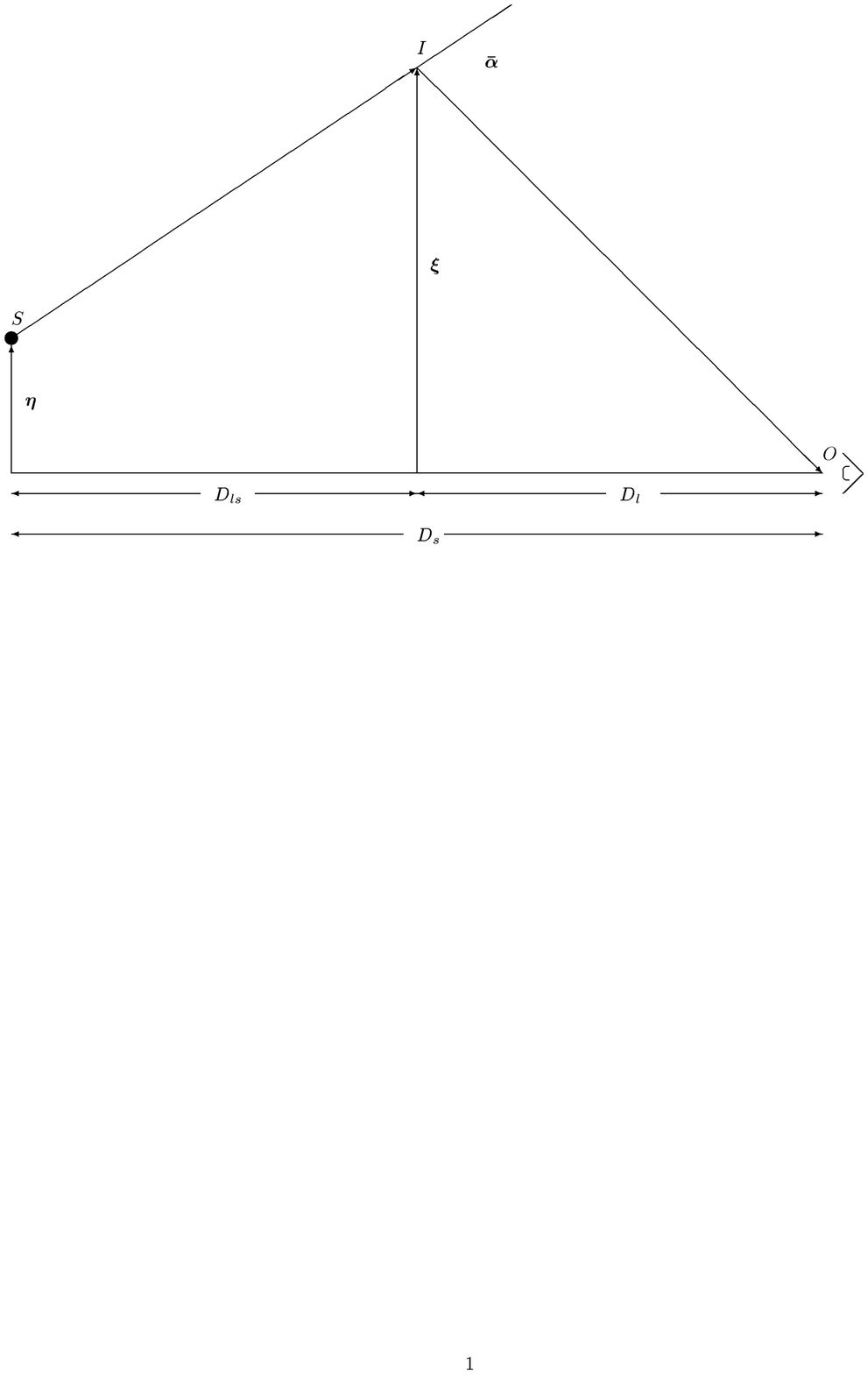}
\end{figure}

Before we consider the specific example of a long cosmic string lens,
let us first take a detour into the basic theory of gravitational
lensing.  For most calculations, it is more than adequate to consider
photon trajectories as the path taken by null geodesics like those
discussed in the previous section, so that the lensing system is
described by simple geometry and wave properties are ignored.  In
figure \ref{lensdiag} we show a pictorial representation of a
gravitational lensing system with an optical axis defined to roughly
intersect the center of the lens.  A photon is emitted from a source
$S$ displaced from the optical axis by a vector $\bbox{\eta}$ at a
distance of $D_s$ from an observer and $D_{ls}$ from the lens, and it
travels in a straight line until it reaches the plane of the lens.
There, its trajectory is deflected by a vector $\bar{\bbox{\alpha}}$,
defined as the initial trajectory minus the final, and it again
travels in a straight line to the observer located at a distance $D_l$
from the lens.  This approximation is known as the thin lens
approximation because the lens and the deflection it induces are
presumed to occur in a single plane.  To first order, this is valid
for long cosmic strings.  The location of the image $I$, where the ray
intersects the lens plane, defines the vector $\bbox{\xi}$ which is
measured from the the optical axis to $I$ in the lens plane.  Since we
will be considering sources and observers that are separated on
cosmological scales, The distances $D_{l}$, $D_{ls}$ and $D_{s}$ are
angular diameter distances (see eq. (\ref{angulardiameter})).

In the small angle limit, the condition that a ray emitted from the
source will have an image seen at $I$ by an observer at $O$ requires
that 
%
%
\begin{equation}
\label{lens}
\bbox{\eta} = {D_s \over D_l} \bbox{\xi} - D_{ls}
\bar{\bbox{\alpha}}(\bbox{\xi}) .
\end{equation}
This equation, often referred to as the lens equation, gives one the
unique source location for any observed image, but its inverse,
however, is not unique, as a single source may generate more that one
image.  It is convenient to recast the lens equation into a
dimensionless form by dividing through by the length $D_l$, so that
the coordinates are given units of angular displacement with respect
to the optical axis.  Thus we define a new set of variables
%
\begin{eqnarray}
\label{dimensonlessvar}
\bbox{x} &\equiv& {\bbox{\xi} \over D_l}, \\ \nonumber
\bbox{y} &\equiv& {\bbox{\eta} \over D_s}, \\ \nonumber
\bbox{\alpha} &\equiv& \bar{\bbox{\alpha}} {D_{ls}\over D_s}, \\ 
\end{eqnarray}
which yield a dimensionless lens equation
%
%
\begin{equation}
\label{lens}
\bbox{y} = \bbox{x} - \bbox{\alpha}(\bbox{x}).
\end{equation}

The above equation can be inverted to give the location of images for
a particular source location, but one can also use the information
contained therein to determine the magnification of those images.
Suppose a source emits a narrow pencil beam of photons which subtends a
solid angle $d\Omega^*$, while the image of the of the source beam
subtends an angle $d\Omega$, so that the ratio of the solid angles
$d\Omega^* /d\Omega$ will give the flux magnification of the image to
the source.  Using the lens equation, one can derive the magnification
by considering the Jacobian matrix
%
%
\begin{equation}
\label{jacobian}
A_{ij} = {\partial y_i \over \partial x_j},
\end{equation}
and observing that the magnification factor ${\cal M}(\bbox{x}) =
d\Omega^*(\bbox{x}) /d\Omega(\bbox{x})$ is given by the inverse of the
determinant of $A_{ij}$,
%
\begin{equation}
\label{determinant}
{\cal M}(\bbox{x}) = {1 \over \det A(\bbox{x})}.
\end{equation}
Note that this same factor will give the angular magnification for
extended objects, so gravitational lenses conserve surface brightness. 

\section{Numerical Calculations with Long Strings}
\subsection{Constructing Long String Networks}
\label{network}
In \S \ref{deflection}, we derived an expression for the photon
deflection that will be relevant to gravitational lensing with cosmic
strings; now we must consider the problem of generating realistic long
cosmic strings. Analytic work, in particular scaling soultions, has
successufully described some of the average properties of long
strings, but it cannot give the details of the structure of a
particular string. Only by simulating networks of interacting strings
can we hope to generate realistic structure.  This exercise can be
greatly simplified if we restrict ourselves to flat, Minkowski space
rather than considering the expanding universe, but  the resulting
structure of these strings will not quantitatively agree with those
produced by expanding universe models. However, there should be good
qualitative agreement, sufficient for our lensing analysis.  That is,
we expect that the structure of the lensed images should be similar
enough to those which would result from more accurate string
simulations to make general conclusions about string lenses.  

Let us begin by considering the equations of motion for a cosmic
string.  The dynamics of strings are dominated by their tension,
while gravitational effects are suppressed by factors of order $G
\mu$, so for now we can ignore the back reaction, but we will discuss
its effects later.  We need the string location which is described by
a four vector $f^\mu$ with a time like component $f^0 = t$ and spatial
displacement $\f(\sigma,t)$ that we used in the previous section.
Then, considering only tension, the equations of motion for the string
are
%
%
\begin{equation}
\label{eqofmotion}
\ddot{f}^\mu - f''^{\mu} = 0,
\end{equation}
where dots refer to derivatives with respect to time and primes refer
to derivatives with respect to $\sigma$.  Our choice of the harmonic
gauge enforces two constraints,
%
%
\begin{equation}
\label{constraint}
\dot{f}^\mu f'_\mu = 0
\end{equation}
and
%
%
\begin{equation}
\label{constraint1}
\dot{f}^2 + f'^2 = 0,
\end{equation}
which restrict the motion to transverse directions and mandate
conservation of energy respectively.  The stress energy can also be
expressed as a function of $\f$ and has the form
%
%
\begin{eqnarray}
\label{stressenergy}
T_{\mu \nu} = \mu \int d\sigma~(\dot{f}_\mu \dot{f}_\nu -f'_\mu
f'_\nu) \delta^{(3)}(\bbox{x} - \f(\sigma,t)),
\end{eqnarray}
consistent with the form suggested in eq.\ (\ref{stress}).  

For non--interacting strings, it would be sufficient to specify some
initial conditions consistent with the gauge constraints and evolve
them using the wave equation.  However, real strings can interact when
two different segments intersect, causing them to reconnect with the
opposite segment.  To handle both the evolution and intersection of
the string network, we shall turn to a clever algorithm first proposed
by Smith and Villenkin \cite{smith}.  The foundation of the
Smith--Villenkin algorithm is the fact that for a set of points
equally spaced in $\sigma$, separated by $\delta$, the wave equation
(\ref{eqofmotion}) can be reduced exactly to a finite difference
equation on a lattice of $\sigma$ and $t$ points. In terms of the
displacement vector $\bbox{f}$, we get
%
%
\begin{equation}
\label{eqofmotionlattice2nd}
\bbox{f}(\sigma,t+\delta) =\bbox{f}(\sigma+\delta,t) +
\bbox{f}(\sigma-\delta,t) - \bbox{f}(\sigma,t-\delta)
\end{equation}
This second order equation can be reduced to a pair of first order
equations if we consider the velocity, defined as
%
%
\begin{equation}
\label{velocity}
\dot{\bbox{f}} \equiv \left\{ \f(\sigma,t+\delta)- {1 \over 2} \left[
\f(\sigma+\delta,t) + \f(\sigma-\delta,t)\right] \right\}/\delta.
\end{equation}
Thus, we get
%
%
\begin{equation}
\label{feq}
\f(\sigma,t+\delta) = {1\over 2}
[\f(\sigma+\delta,t)+\f(\sigma-\delta,t)] + \dot{\f}(\sigma,t)\delta,
\end{equation}
and
%
%
\begin{equation}
\label{fdoteq}
\dot{\f}(\sigma,t+\delta) = {1 \over 2} [\dot{\f}(\sigma+\delta,t) +
\dot{\f}(\sigma-\delta,t)] + [\f(\sigma+2\delta,t) -2\f(\sigma,t) +
\f(\sigma-2\delta,t)]/4\delta.
\end{equation}
A complete solution can be specified by initially fixing the positions
for every even point on the $\sigma$ lattice and velocities for every
odd point.  After the first time step, eq.'s (\ref{feq} \&
\ref{fdoteq}) will give the positions for each odd point on the
$\sigma$ lattice and velocities for each even point on the
lattice. After the second time step, even points will again be
positions and odd points will again be velocities, and so on, so that
in general, the plane of $\sigma$ and $t$ will be filled with
interlocking diamond lattices of positions and velocities. 

The next challenge is to satisfy the gauge constraints on the lattice.
In eq. (\ref{velocity}) we have given a discrete velocity, and now we
need a discrete version for the $\sigma$ derivative.  It has the
obvious form
%
%
\begin{equation}
\label{fprime}
\f'(\sigma,t) \equiv
[\f(\sigma+\delta,t)-\f(\sigma-\delta,t)]/2\delta,
\end{equation}
where the gauge constraints for the discrete $\f$ remain unchanged,
%
%
\begin{equation}
\label{descretegauge1}
\dot{\f}\cdot \f' = 0,
\end{equation}
and
%
%
\begin{equation}
\label{descretegauge2}
\dot{\f}^2 +  \f'^2 = 1.
\end{equation}
To ensure that these conditions would be preserved through the entire
evolution of the string, Smith and Villenkin proposed discritizing
space in the same way as $\sigma$ and $t$, that is, the space itself
is a lattice of points with spacing $\delta$.  String configurations
are then described by a series of connected links between successive
string points on the $\sigma$ lattice consisting of three possible
types.  The first type is a static link with $\dot{\f} = 0$ and
$|\Delta \f = 2|$, implying that one coordinate of $\Delta \f$ is $\pm
2\delta$ while the others are zero.  The second type is a moving link
for which $|\f'| = |\dot{\f}| = \sqrt{2}/2$.  Two components of
$\Delta \f$ are $\pm \delta$ while the final is zero, and the velocity
$\dot{\f}$ also has two non zero components which are each $\pm 1/2$
and must be normal to $\Delta \f$.  The last type of link is a cusp
where $\Delta \f = 0$.  The velocity of a cusp is 1 corresponding to a
link traveling at the speed of light parallel to one of the axes.  One
can verify easily that all of these links satisfy the gauge conditions
and that the constraints will be preserved by the equations of motion.

Finally, to accurately evolve a string network, we must account for
string inter--commutations which occur when different segments of the
string collide. We have defined links which are connections of two
successive string positions $\f(\sigma,t)$ and $\f(\sigma+\delta,t)$.
A collision occurs when two string points $\f(\sigma_i,t)$ and
$\f(\sigma_j,t)$ both fall on the same location.  If this happens, we
inter--commute the links so now the point $\f(\sigma_i,t)$ is
connected to $\f(\sigma_j+\delta,t)$ and the point $\f(\sigma_j,t)$ is
connected to $\f(\sigma_i+\delta,t)$ with care taken to ensure that
the proper velocity is assigned to each link.  Prior to each time
step, one tests each of the points to see if it lies in the same
position as any other.  One inter--commutes all the colliding links,
and then evolves time for one step $\delta$ and checks the strings
again for inter--commutations.  To ensure that cusps are not mistaken
for collisions, we set a minimum separation in $\sigma$ space,
4$\delta$, required before an inter-commutation on a distinct string
is allowed.  This fixes a minimum loop size, but the structure of the
long strings is insensitive to the choice so long as it is small in
comparison to the overall string length.  In passing we mention that
the order $N^2$ process of testing each point for intersection can be
reduced to an order few $N\log N$ process if one sorts the points with
respect to their positions using a quick sorting routine first.

The Smith--Villenkin algorithm is a powerful tool for evolving
networks of cosmic strings, and now we shall consider the initial
conditions which we will use to produce long string segments for
gravitational lensing.  We start with the standard initial conditions
introduced by Vachaspati and Villenkin\cite{vachaspati}, who lay down
a periodic lattice of phases, and locate strings through the faces of
the lattice cubes which have non--zero winding number.  The result is
a network of strings made of static segments which are parallel to one
of the axes of the box, ideally suited as initial conditions for the
Smith--Villenkin evolution algorithm.  To ensure that space
discritization effects do not strongly influence the resulting
strings, each of these segments was subdivided into sixteen static
links 2$\delta$ long, where the overall segment length is
32$\delta$. Tests of our simulations found that subdividing the
segments with a larger number of links did not change structure of the
evolved network, so spatial discritization effects have been
minimized. Ideally, we would like to evolve this network until it has
completely relaxed, and use the resulting long strings for our lensing
calculations. Unfortunately, the periodic boundary conditions
insure that there can be no net string flux through the box, meaning
that given sufficient time, all of the long strings will fragment into
loops leaving us nothing to study.  To minimize these periodic
effects, we evolve the network for a time equal to half the light
crossing, {\it i.e.} if the box is $n\delta$ wide, we take $n/2$ time
steps, each $\delta$ long. We expect scales on the box size to
preserve the structure of the initial conditions, but on scales
smaller than $n \delta / 2$, interactions should have sufficient time
to relax the system to its final state.  

\begin{figure}
\caption{\protect\label{fractal}
String conformal length $L$ plotted as a function of point separation
distance $D$. The solid curve is for $t=0$, the dotted for
$t=256\delta$ and the dashed for $t= 512\delta$.  The periodic length
of the box is $1024\delta$.}
\epsfxsize = \hsize \epsfbox{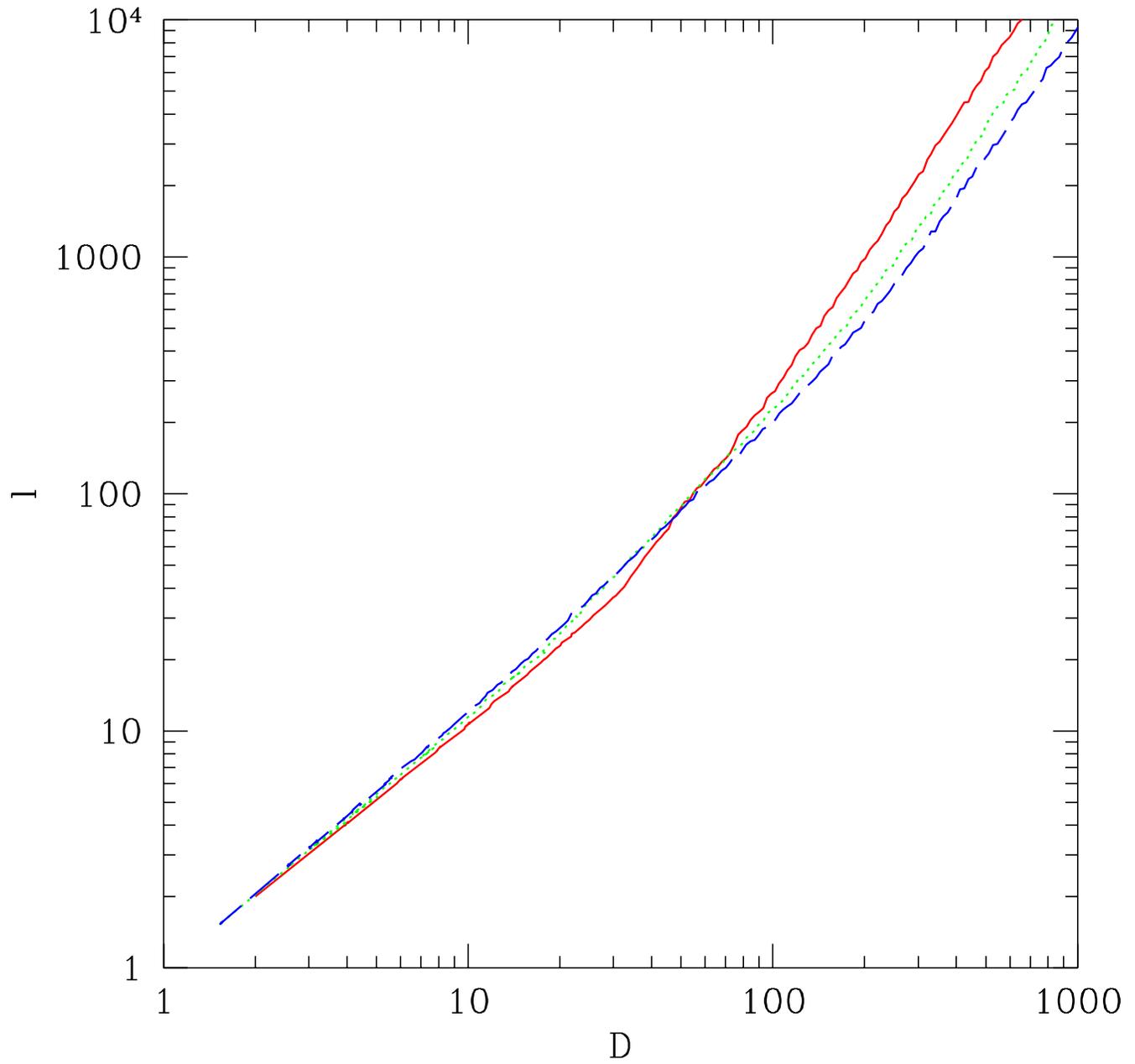}
\end{figure}

Previous analysis of evolved string networks has shown that the small
scale structure of the surviving long strings is self--similar, that
is, the structure is statistically the same on all scales well below
the horizon\cite{refs}. We compare our results with our predecessors,
as a consistency check.  The simplest test is to measure the fractal
dimension of the string defined as the exponent $n$ such that $L
\propto D^n$, where $L$ is the mean conformal length of string
measured between points separated by a distance $D$ in physical space.
For self--similar structures, $n$ is a constant for all $D$, and for
the particular example of a random walk string, $n = 2.0$.  In figure
\ref{fractal} we show a log--log graph of the conformal length $L$
plotted as a function of $D$ for the long strings produced in a box
simulation with periodic length 1024$\delta$ at times $t =
0\delta,~256\delta$, and $512\delta$.  Note that our initial segments
were formed in a $32^3$ box of phases and each initial segment was
32$\delta$.  We can see that for the unevolved strings that the
fractal dimension is roughly uniform above the initial link scale.  A
linear regression of these points gives a slope of $n = 2.0$ which is
what one expects from a random walk.  As the simulation evolves, we
see that the shorter length scales begin to relax into a new structure
with a smaller fractal dimension, and by the time $t = 512\delta$,
scale below $D \sim 512\delta$ have almost completely relaxed. A fit
to these points gives a fractal dimension of $n = 1.3$, a result
consistent with those of Sakellariadou and Villenkin
\cite{sakellariadou}, and interestingly, also consistent with the
early time results of expanding universe simulations.  In this last
type of simulation, apparently the strings first fragment and relax on
time scales that are short compared to the expansion time of the
universes and then are stretched so that $n$ falls below the initial
flat space value.

\subsection{Finding Images With Long Strings}
\label{images}
Now that we have a network of realistic long strings to work with, we
consider how to use them in gravitational lensing systems.  The best
approach would be to evolve a large network with very fine spatial
resolution (on the order of $10^{-5}$ the box length, where we could
associate that scale with a few times the horizon scale), and just
send photons through the box.  Of course, the numerical resolution
required to perform such a calculation is well beyond the capacity of
current computers, so as an alternative, we exploit the fractal nature
of the small scale structure of the string discussed in the previous
subsection.  Sections of string which can fit in a box of length about
half the simulation box relax into a self similar structure described
by a constant fractal dimension, meaning that if we were to magnify a
small piece of this string, we would observe the same structure in
proportion to the new scale.  Thus, to use our long strings which are
not resolved on the scales relevant to gravitational lensing, we need
only rescale the string, as long as we restrict ourselves to segments
which have had sufficient time to relax.

Let us be more specific about precisely how we accomplish this.  From
our simulations described in the previous section, we have a periodic
box filled with loops and long strings.  We can eliminate the loops by
considering only strings which are significantly longer than the box
length, leaving the long strings which tend to wrap around the box
periodically once or more.  It is the smaller scales, for which the
string has relaxed to its final structure, that we wish to consider.
To do so, we remove a long string from the box entirely, laying it
end to end no longer in a periodic box.  If the string wraps around
the original network box more than once, we connect the the box size
segments by shifting the endpoints to make one super--long string.
Now we need to determine the the light cone projection of the string
required to find the deflection given by eq.\ (\ref{deflecttilde}).
We accomplish this by allowing our single long string to evolve
independently of the network, turning off all inter--commutations and
connecting the end points periodically. Inter--commutations are
ignored because we do not want to alter the small structure of the
string.  A photon is presumed to travel along one of the axes, and the
location of the intersection of each of the string points with the
light cone, along with the string velocity is recorded.  Since each
string point is equal spaced in $\sigma$, we can use the set of light
cone projected points to reduce the integral in eq.\
(\ref{deflecttilde}) to a discrete sum.  We do not, however, wish to
use the entire long string, since the structure has not relaxed on the
largest scales.  Instead, starting at an arbitrary point, we select
shorter segments--those which can fit in a box with side length
half that of the original network--and use only these points to
calculate the photon deflection. 

In truncating the summation and considering only a finite string
segment, we obviously introduce errors which we would like to
quantify. As a order of magnitude estimate, let us consider the
special case of an finite straight string segment perpendicular to the
photon trajectory with equal lengths $\ell$ above and below the photon
axis. We consider the photon to be moving along the $z$ axis and place
the string segment parallel to the $y$ axis a distance $d$ from the
origin (defined by the photon) so that the deflection will be along
the $x$ direction.  Using eq.\ (\ref{deflect}), while observing that
$f_\| = 0$ and $F_{\mu\nu} \gamma^\mu\gamma^\nu = 1$, we see that the
magnitude of the deflection will be proportional to
%
%
\begin{equation}
\label{truncate}
\int_{-\ell}^\ell { d \over \sigma^2 +d^2} = 2 \tan^{-1}(\ell/d),
\end{equation}
and the truncation error, that is the difference between $\ell
\rightarrow \infty$ and finite $\ell$, goes as $\pi - 2 \tan^{-1}(\ell/d)$.
In the limit of large $\ell$ this is approximately $2 d / \ell$.  If
instead we wish to look at a fractal string, then we should replace the
$\sigma^2$ in the integral with something proportional to
$\sigma^{2/n}$ which gives a truncation error which falls like $(d /
\ell)^{(2 / n) - 1}$.  So, we see that the error depends on how far
the photon passes from the string, which is fortuitous since most of
the interesting lensing occurs for photons which pass close to the
string segment.  We can also see that the thin lens approximation is
justified.  So long as $d$ is much smaller than the distances of the
source and observer from the lens,$\sim D$, then the deflection will
occur in a region within a few lengths $d$ from the string which, and
including the effects arising from the non--instantaneous deflection
would be second order in $d/D$.

Given the projected string segment, we have everything necessary to
examine gravitational lensing, but now the challenge is to solve the
lens equation.  Currently there is no fast way to invert a
multidimensional equation like eq. (\ref{lens}), so our only choice is
to solve the problem by brute force.  Given a source and lens
redshift, we calculate the deflection $\bbox{\alpha}(\bbox{x})$ on a
uniform grid in $\bbox{x}$ space, and then use the lens equation to
solve for $\bbox{y}$ at each $\bbox{x}$. In other words, we are
mapping a uniform grid in the image plane back onto the source plane
where we can use this map to locate the images of a particular source.
We compare triangles made of nearest neighbors in the image grid
mapped onto the source plane to the locations of the sources.  If a
source point is enclosed by a mapped image triangle, then its image
must lie somewhere in that triangle in the image plane, and thus, for
any source point, we can locate its image to the uncertainty given by
the image grid spacing.  One must be careful, though, when considering
image triangles that enclose a piece of string because the photon
deflection is discontinuous as $\bbox{x}$ crosses the string. Images
in these triangles should be ignored because they represent photons
which must pass through the string itself to be observed and are
therefore spurious.  Since our strings are resolved down to the scale
$\delta$, it is natural that the image grid spacing should be $\delta$
as well.  And, having assumed a self--similar structure for the
string, we are free to choose the physical scale that $\delta$
represents.  For the objects we shall consider for lensing, we find
that a good choice for this scale is $\delta / D_l = 0.1$ arc sec.

\begin{figure}
\caption{\protect\label{quasar1}
A string--quasar lensing system.  The string is shown as a dashed line
while the source is shown as a hatched circle with images shown as
open circles.  The relative areas are indicative of the magnifications
of the images.}
\epsfxsize = \hsize \epsfbox{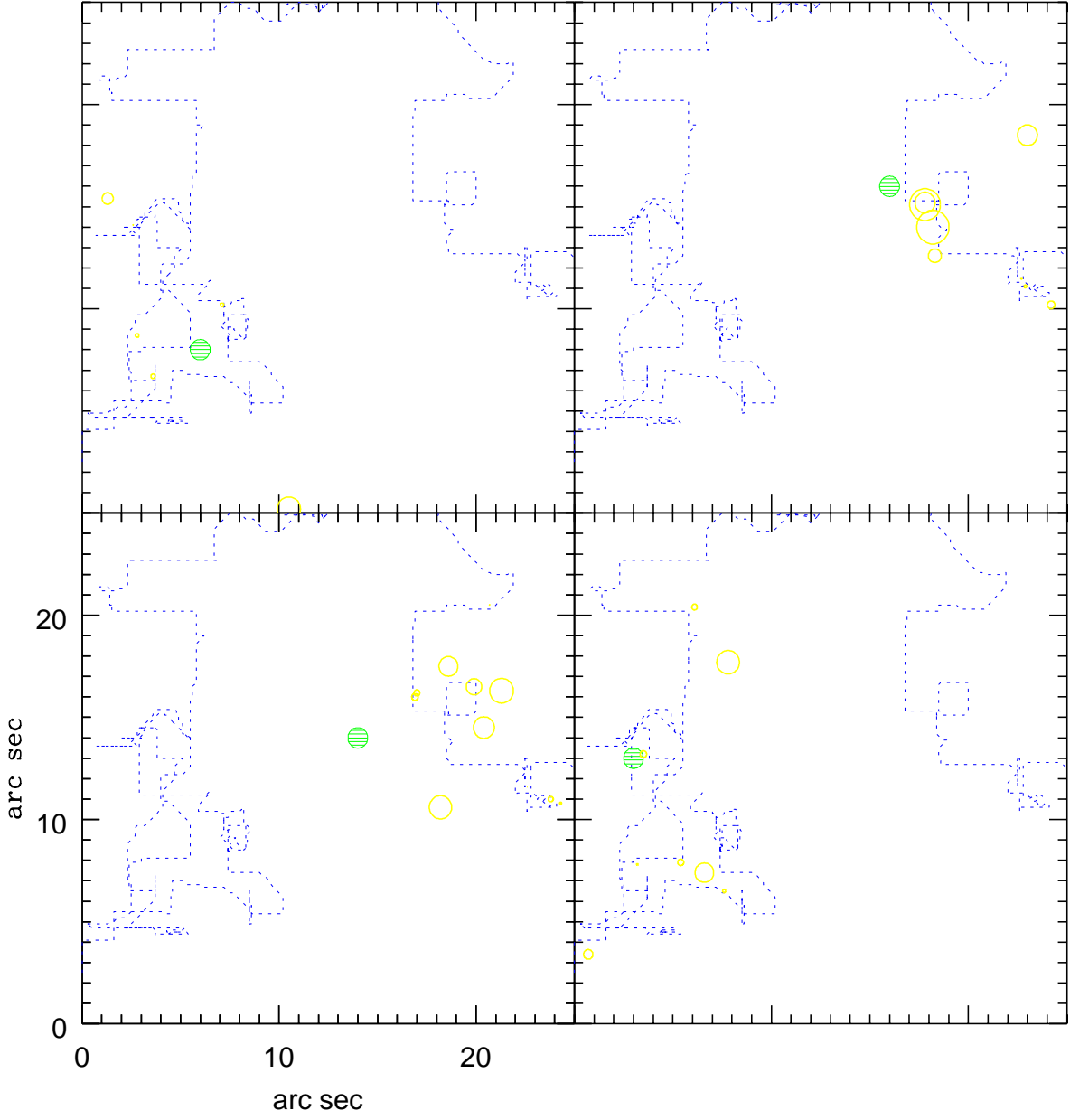}
\end{figure}

\begin{figure}
\caption{\protect\label{quasar2}
Same as Fig.\ \protect\ref{quasar1}}
\epsfxsize = \hsize \epsfbox{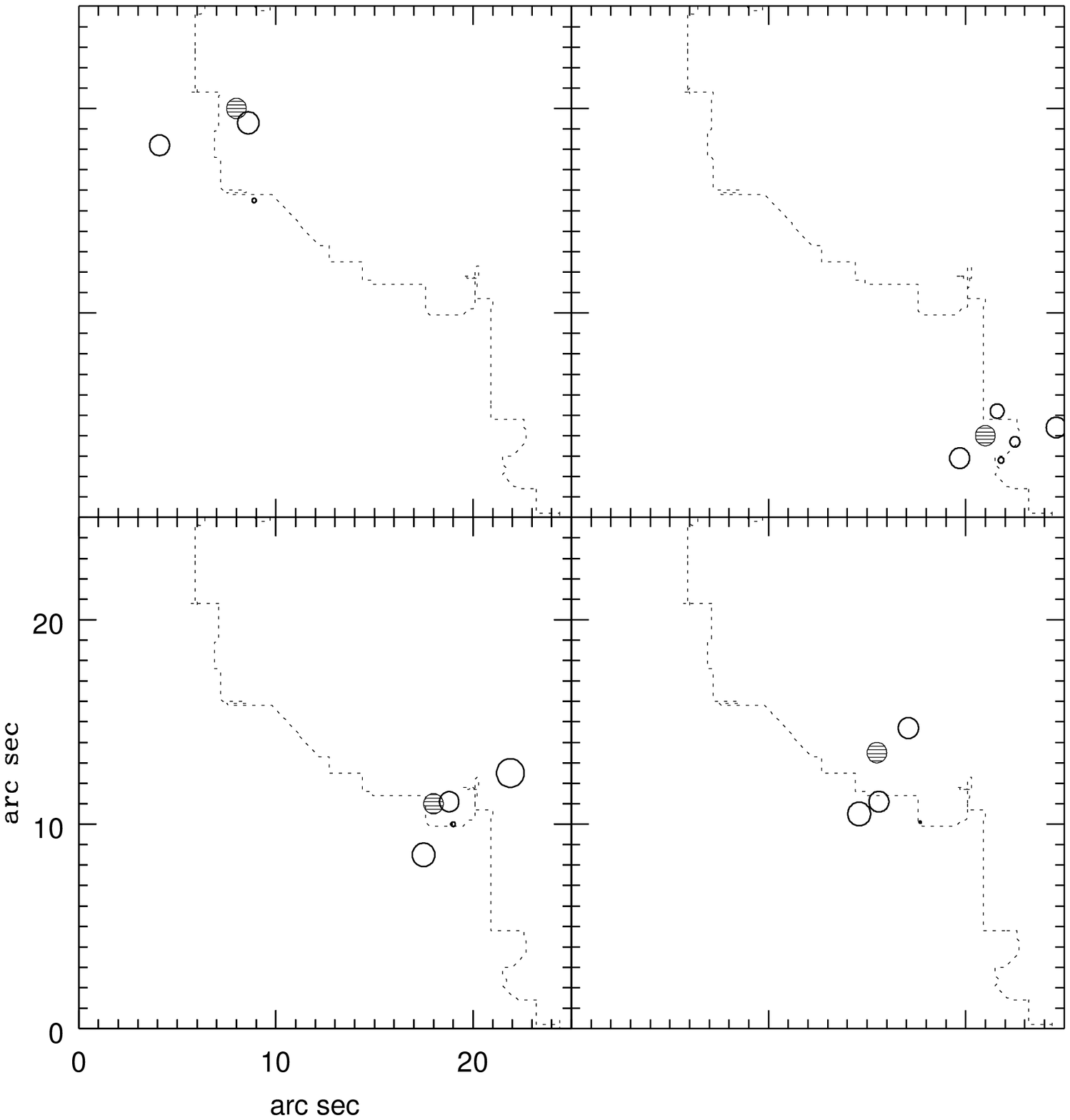}
\end{figure}

\begin{figure}
\caption{\protect\label{quasar3}
Same as Fig.\ \protect\ref{quasar1}}
\epsfxsize = \hsize \epsfbox{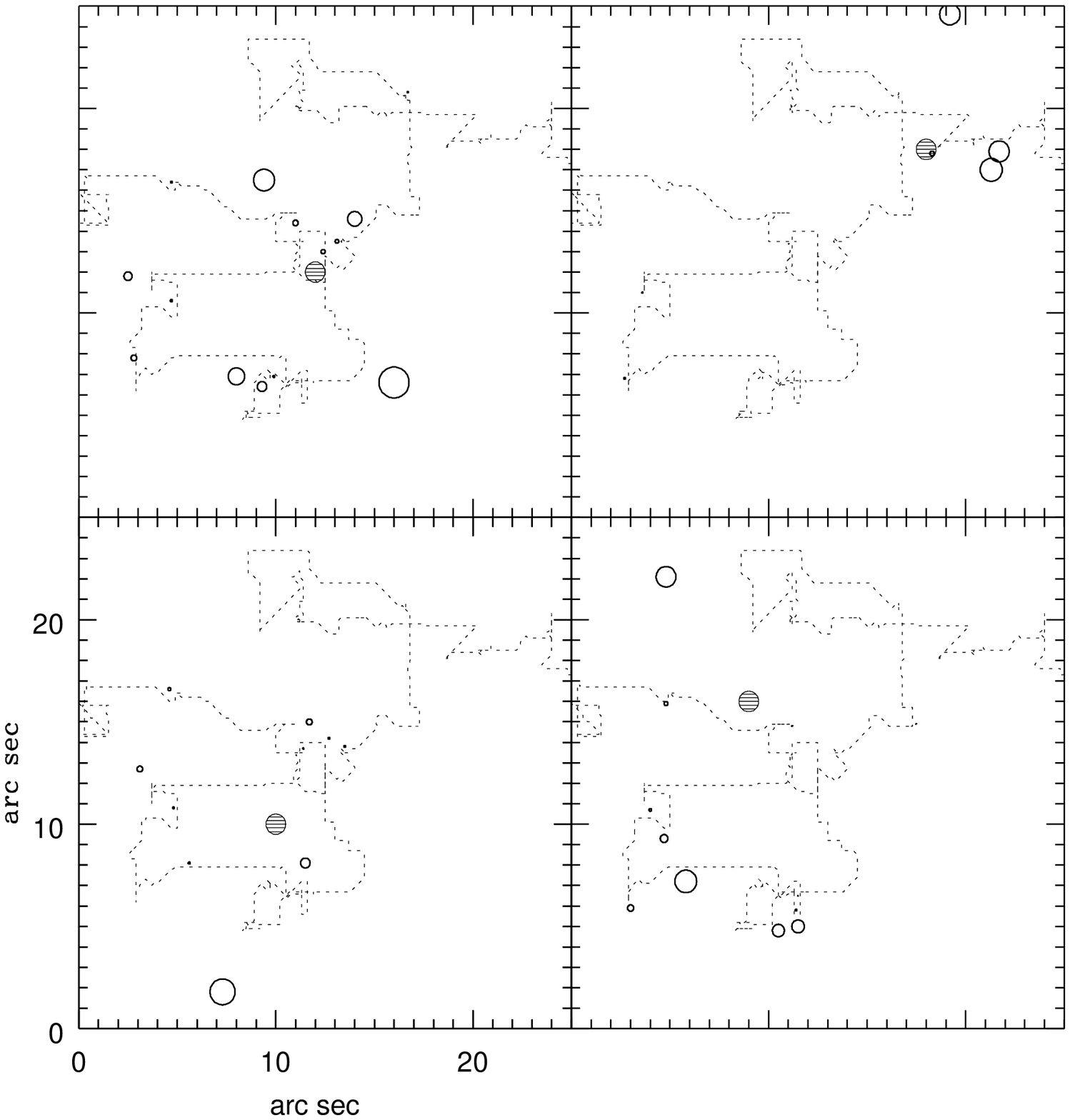}
\end{figure}

\begin{figure}
\caption{\protect\label{source1}
A galaxy--string lensing system.  Shown is the full string used in
calculating the photon deflection along with several galactic sources.
The dashed box shows the region in which the images of these sources
are calculated.  The string is located at $z=1$ and the galaxies at $z=2$.}
\epsfxsize = \hsize \epsfbox{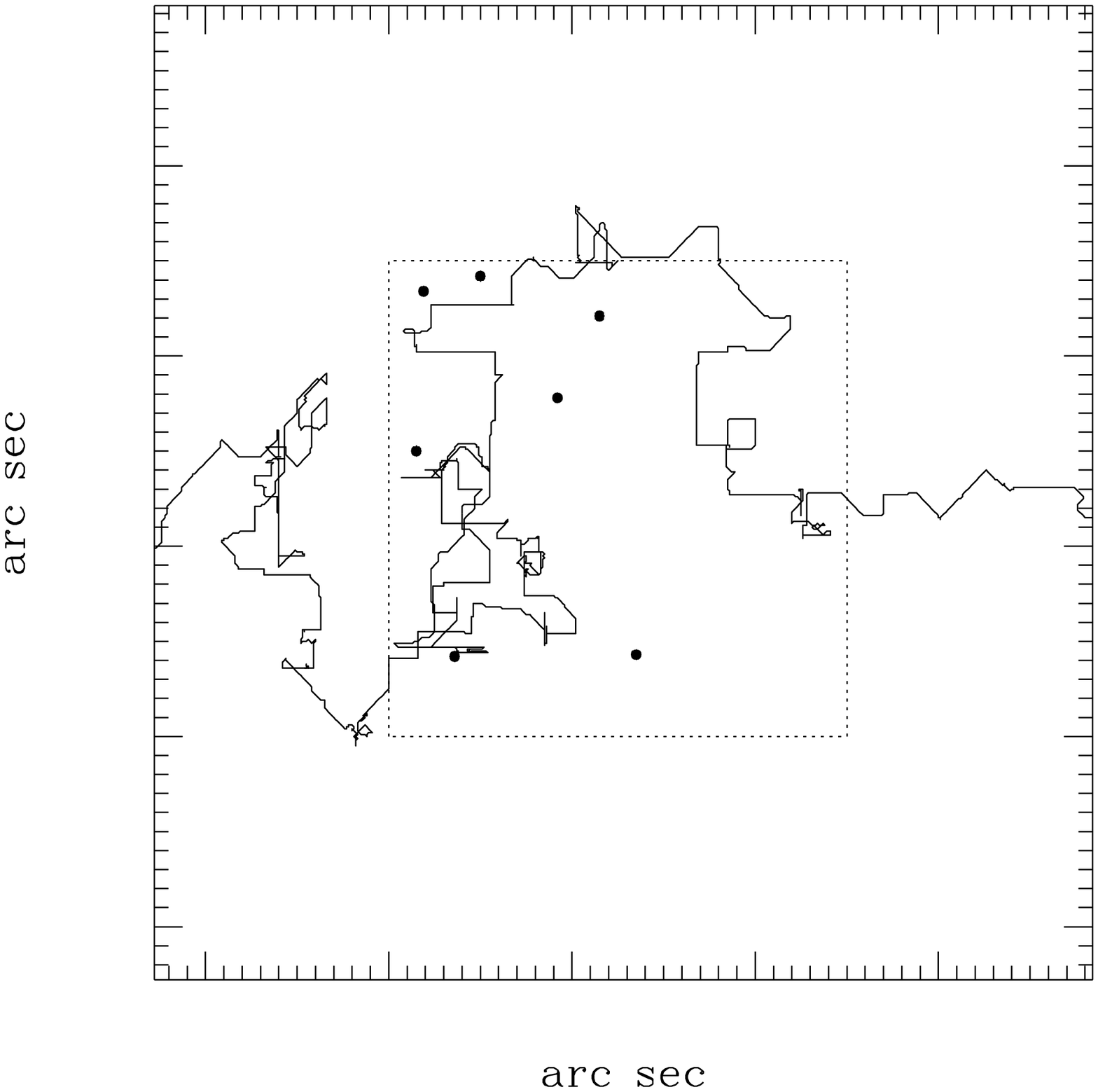}
\end{figure}

\begin{figure}
\caption{\protect\label{source2}
Another string--galaxy system like Fig.\ \protect\ref{source1}}
\epsfxsize = \hsize \epsfbox{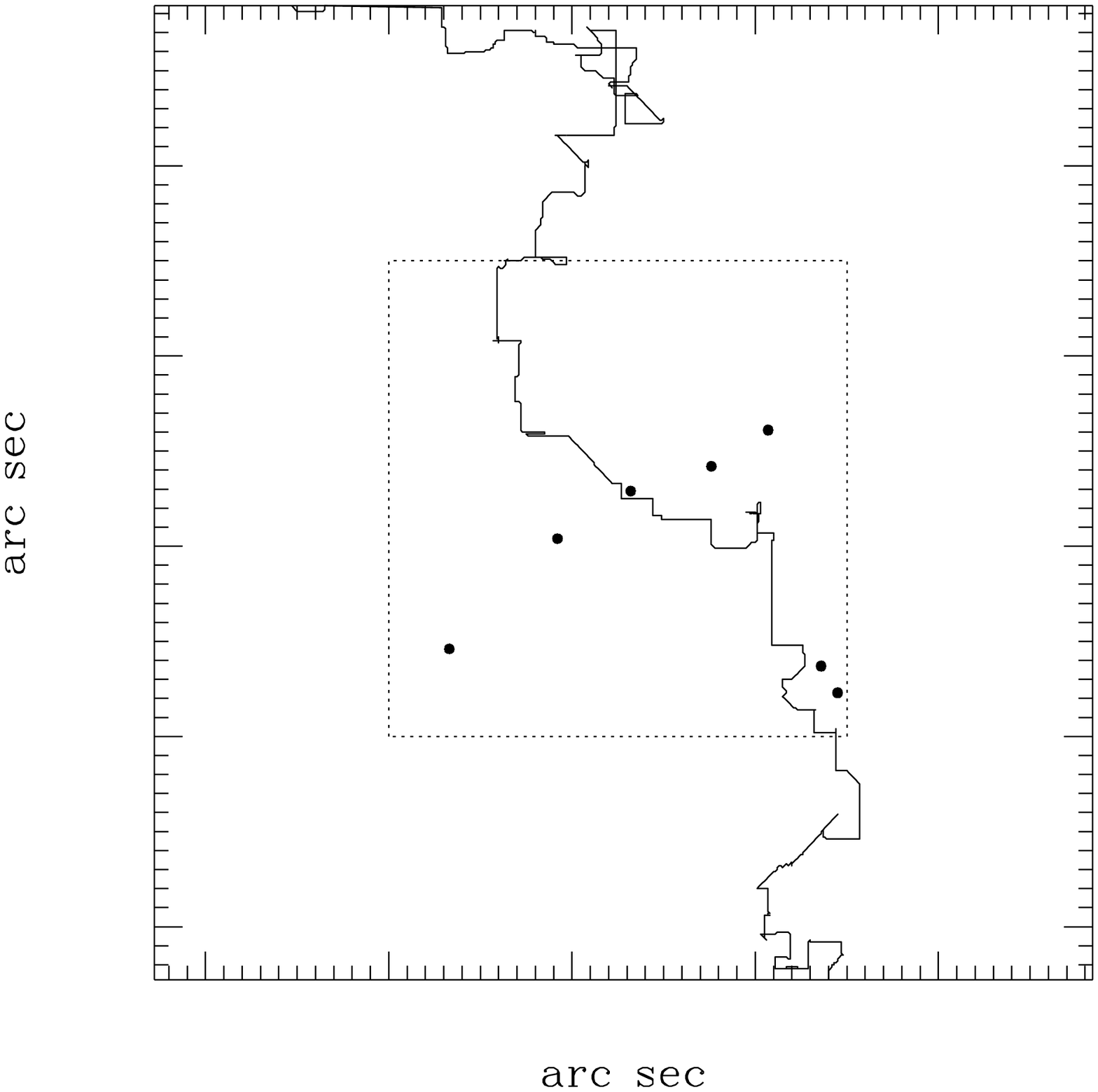}
\end{figure}

\begin{figure}
\caption{\protect\label{source3}
Another string--galaxy system like Fig.\ \protect\ref{source1}}
\epsfxsize = \hsize \epsfbox{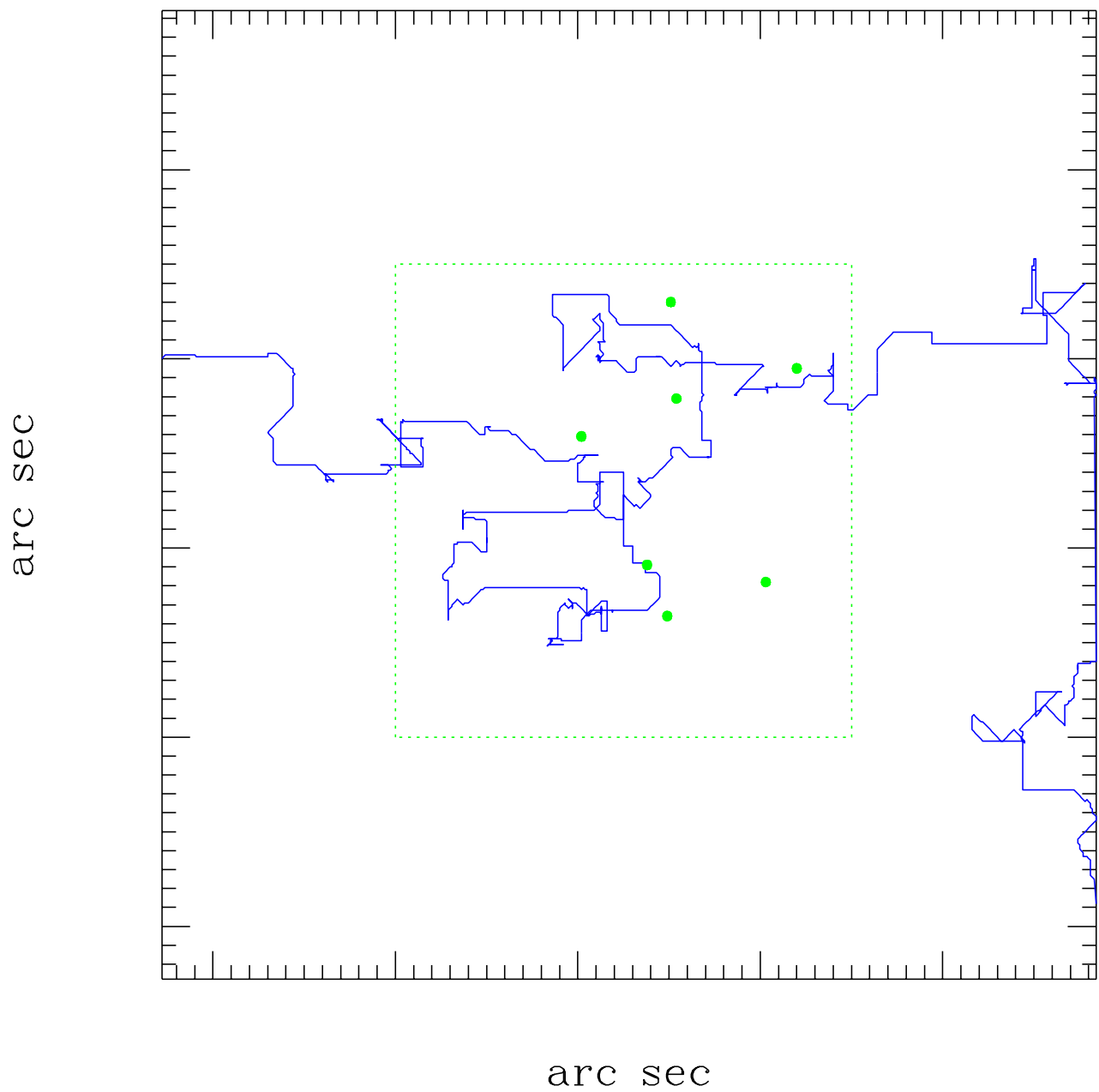}
\end{figure}

Using the techniques described above, we shall consider two types of
sources, point--like and extended, which will roughly correspond to
quasars and galaxies respectively.  Let us first consider quasars as
they represent what are likely the best objects to look at when trying
to observe strings through lensing.  Quasars possess a trio of virtues
regarding lensing, namely they are high redshift objects ($z \sim
1-5$), they are bright and therefore easily observed, and they are
typically separated by large enough distances that the chance
coincidence of two different quasars being separated at the same
scales as typical lensing systems is rare. However, there are a
sufficient number to make the observation of string quasar lensing
systems probable (see \S \ref{probability}).  Quasars are compact,
distant, and cannot be spatially resolved with current
technology. Thus we shall treat them as idealized point sources.  The
locations of the resulting images are found as we have described,
along with their magnifications which are determined by eq.\
(\ref{determinant}).  Figures \ref{quasar1}, \ref{quasar2}, and
\ref{quasar3} show the results quasar lensing with three different
segments of cosmic string.  In each panel, the source location is
shown as a hatched circle, the images as open circles, and the
projected string segment as a dashed line.  The relative area of the
image circles gives the ratios of magnifications of the image to the
source.  The string is located at a redshift of one and the quasar is
located at a redshift of two, while the angular size of each
panel is 25 arc sec.  The full length of string used in the lensing
calculations is not shown, but we use the same segments for the galaxy
lenses, and we do show the full segment in figures \ref{source1},
\ref{source2}, and \ref{source3} respectively. Typical in many of
these examples are a number of small demagnified images which reside
close to the string itself.  These results are qualitatively similar
to those one would see if we replaced the string by several point
masses with masses similar to the energy in the string.  With point
masses, The deflection induced by neighboring points cancels around
the midpoint between the two masses.  Small images tend to form here
because the deflection angle changes rapidly around the minima.  For
strings, it is the kinks and wiggles which provide a similar
opportunity.  Inside a kink, contributions from different parts of the
string can cancel, leading to rapid changes in the deflection.  Kinks
can also produce small images by just being large concentrations of
energy.  They produce images in a manner similar to point masses when
the source is outside the Einstein ring.  The secondary image is smaller
and therefore demagnified.

\begin{figure}
\caption{\protect\label{image1}
The images observed for the lensing system shown in Fig.\
\protect\ref{source1}}
\epsfxsize = \hsize \epsfbox{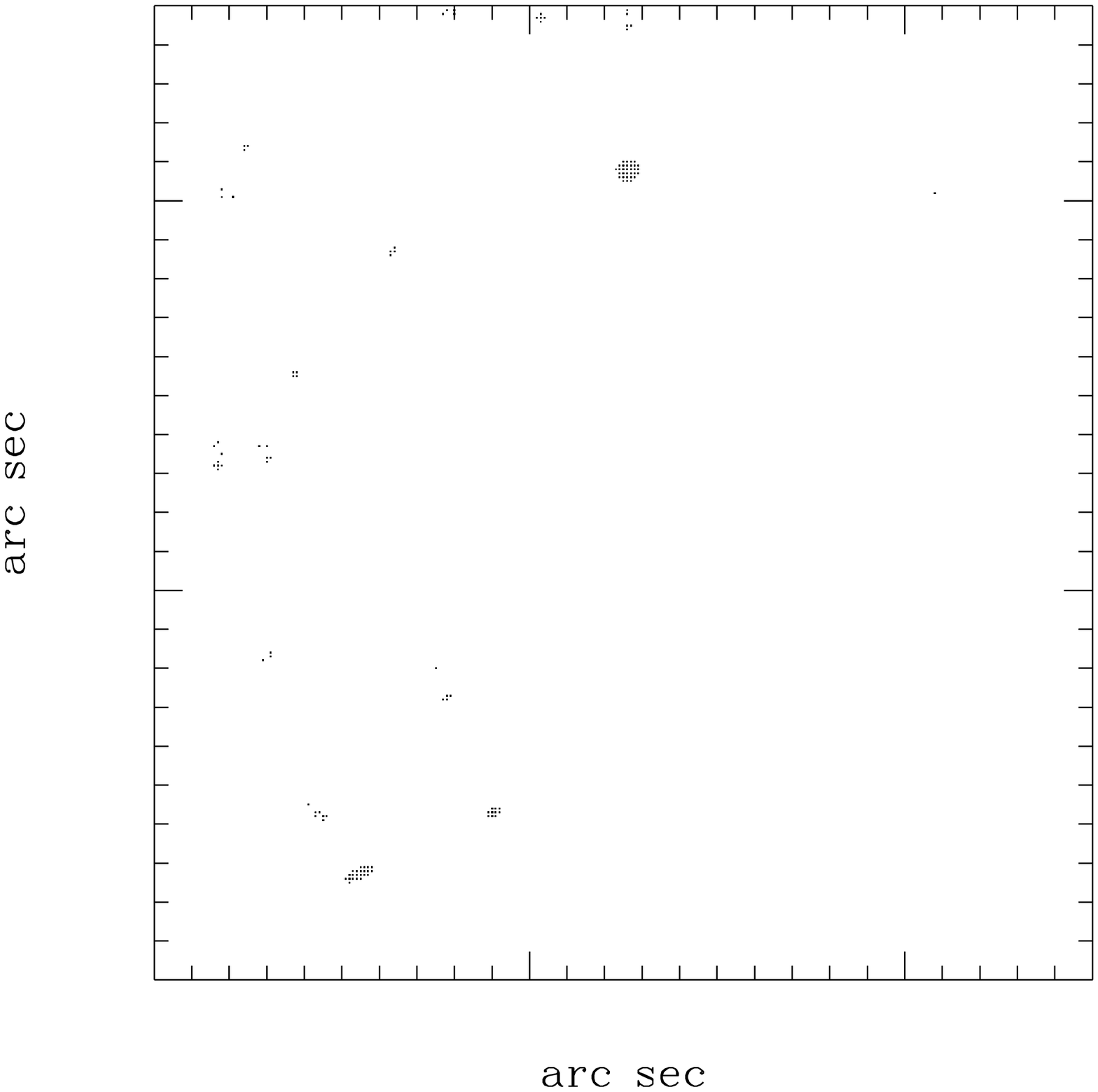}
\end{figure}

\begin{figure}
\caption{\protect\label{image2}
The images observed for the lensing system shown in Fig.\
\protect\ref{source2}}
\epsfxsize = \hsize \epsfbox{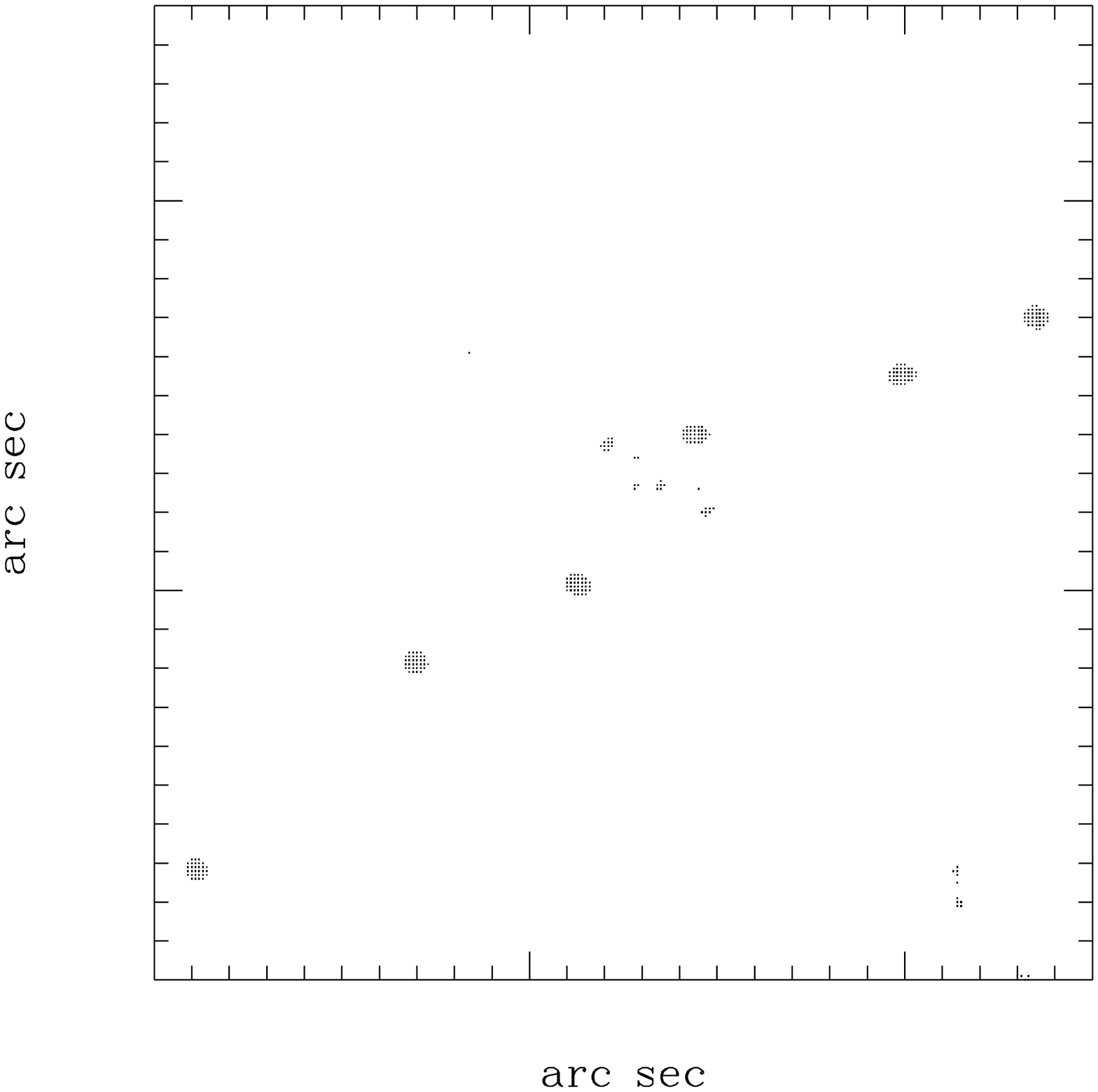}
\end{figure}

\begin{figure}
\caption{\protect\label{image3}
The images observed for the lensing system shown in Fig.\
\protect\ref{source3}}
\epsfxsize = \hsize \epsfbox{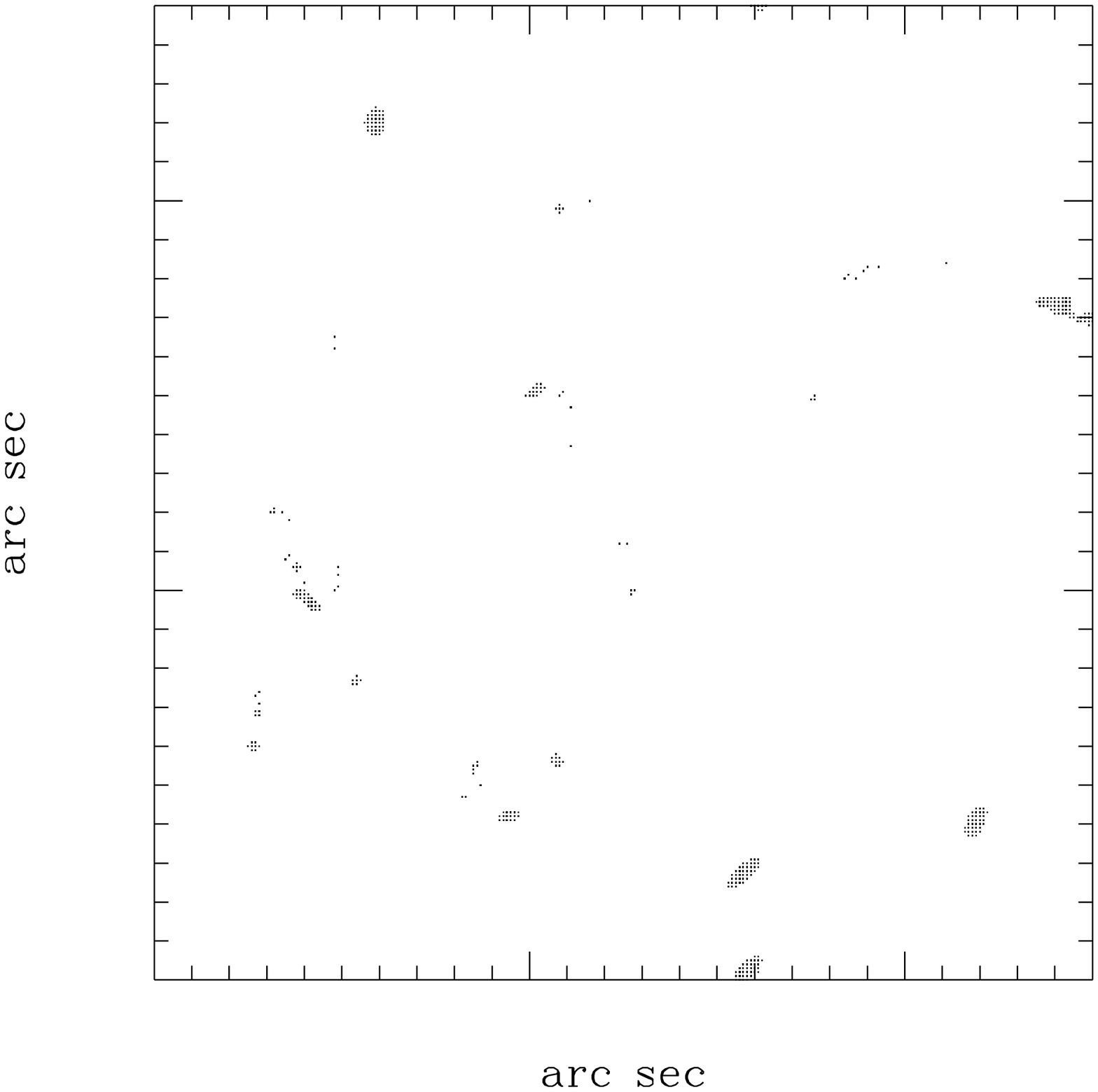}
\end{figure}

High redshift galaxies also provide interesting candidates for cosmic
string lensing, but observing these systems is significantly more
challenging.  The greatest difficulty lies in observing such objects
because they are so faint.  Things are further complicated because
foreground galaxies may also contaminate the systems making it more
difficult to observe the images. Also, typical galaxies are not
spherical making it difficult to determine if one is observing a
lensing arc or merely an edge on galaxy.  However, to qualitatively
illustrate a galaxy lensing system, we consider an idealized case of a
set of extended spherical sources located at a redshift of $z=2$,
lensed by a string at $z=1$.  The angular size of the sources is 0.5
arc sec which corresponds roughly to the observed size of high
redshift galaxies.  Seven such sources are scattered randomly in side
a $(25~{\rm arc~ sec})^2$ viewing area mimicking the approximate
angular density of real high redshift galaxies \cite{sawicki}.  In
figures \ref{source1},
\ref{source2}, and \ref{source3} we show the full string segment used
in the lensing calculation along with the sources to be lensed.  The
dashed box shows the area for which we calculate the images, where
outside this region, we expect that the photon deflections will not be
particularly accurate due to truncation error.  In figures  \ref{image1},
\ref{image2}, and \ref{image3} we show the observed images
corresponding their respective sources, and again we note the
proliferation of smaller, demagnified images similar to those observed
in the quasar images.
 
\section{Conclusions}
\label{conclusion}
From the results in the previous section, we can see that cosmic
strings can produce images which have characteristics unlike the more
prosaic galaxy lens.  In particular, the proliferation of small
demagnified images is a signature that may be unique to long cosmic
strings.  This suggest the exciting possibility that a cosmic string
can be positively identified through gravitational lensing, confirming
the existence of these topological defects.  Unfortunately, our
enthusiasm must be tempered by two important caveats.  The first is
that these strings are the product of Minkowski space simulations and
do not include the effects of the expanding universe.  When theses are
considered, the structure on the string is stretched so that the
fractal dimension falls to approximately $n = 1.1$ \cite{bouchet}, so
we expect real strings to have smaller kinks and wiggles.  The other
effect that we ignored is that of gravitational radiation.  For loops
it is well known that they radiate energy at a rate of $\Gamma G \mu$
where $\Gamma$ is on the order of 100, so loops shorter than $\Gamma G
\mu t$ will have radiated completely away. The effects of
gravitational rediation on the structure of fractal strings is not as
well understood.  An alalytic example of a helical string has been
calculated by Sakellariadou \cite{sakellariadou1} while the case of
small amplitude kinks has been considered by Hindmarsh
\cite{hindmarsh}. Their results suggest that fluctuations on long
strings should have a life expectancy of about $d/G\mu$ where $d$ is
the typical separation length between kinks--on the same order as the
fluctuation.  So, one might expect that the strings are straight on
scales smaller than $G\mu$ times the horizon scale at the epoch of
lensing. For a string located at $z = 1$ this corresponds to an
angular scale of 0.5 arc sec, so gravitational radiation may just
influence the long string lensing structure.  Our examples then
represent the most extreme results that one should expect, and real
signatures may be less distinct.  However, because the small scale
structure of our strings is responsible for the demagnified images, we
can conclude that gravitational lensing may be a good way to probe
that small scale structure.  If strings are relatively smooth on the
scales we considered in our examples, then any strong lenses observed
should produce a pair of undistorted images like those resulting from
an infinite straight string.  Conversely, if there is significant
small scale structure, then one expects to see a number of demagnified
images like those seen in the figures.

In \S \ref{probability}, we found that quasar--string lensing systems
should constitute about 10\% to 30\% of the observed quasar lenses,
where the rest arise from lensing with galaxies.  We suggest that the
search for gravitational lenses could on its own confirm or rule out
the cosmic string model of structure formation depending on whether
such lenses are observed. Since the strings are obviously correlated,
a large angle of sky coverage with a large number of quasars is
required.  In fact, precisely such a survey is currently being
developed, namely the Sloan Digital Sky Survey.  Some $10^5$ quasars
with a mean redshift of two are to be observed in a $\pi$ steradian
slice of the sky.  We have estimated that one should observe on the
order of ten string lenses in the SDSS for an $\Omega = 1$, cosmic
string model.  The failure to observe any string lensed quasars would
require $G\mu < 10^6$, making strings an unlikely candidate for
structure formation.  Conversely, should a suspected string lens be
observed, it is possible to confirm it by looking at galaxy
observations in the neighborhood of the lens.  With precise
observations, one expects to see lensing of high redshift galaxies
from nearby parts of the same string.  In concert then, quasar
observations followed by galaxy observations could provide definitive
proof of GUT scale cosmic strings, or rule out cosmic strings as a
viable model for the formation of large scale structure.

We would like to thank Tanmay Vachaspati for his help and
suggestions. This work was supported with funding from the Department
of Energy. 


\newpage
%
%
\newpage

\newpage

\newpage

\newpage

\newpage

\newpage


\newpage

\newpage

\newpage

\newpage

\newpage

\newpage

%
%


\begin{references}
\bibitem{allen} B. Allen {\it et al}., Phys. Rev. Lett. {\bf 77}, 3061
(1996)

\bibitem{magajio} J. Magueijo {\it et al}., Phys. Rev. {\bf D54}, 3727
(1996)

\bibitem{bennett} D.P. Bennet and F.R. Bouchet in 
{\it The Formation and Evolution of Cosmic Strings}, eds. 
G.W. Gibbons, S.W. Hawking and T. Vachaspati,
(Cambridge University Press 1990).

\bibitem{turner} E. L. Turner, J. P. Ostriker and J. R. Gott, Astrophys. J.
{\bf 284}, 1 (1984)

\bibitem{white} L. M. Krauss and M. White, Astrophys. J. {\bf 394},
385 (1992) 

\bibitem{delaix} A. A. delaix and T. Vachaspati, Phys. Rev. {\bf D54},
4780 (1996)

\bibitem{smith} A. G. Smith and A. Vilenkin, Phys. Rev. {\bf D36}, 990
(1987).

\bibitem{vachaspati} T. Vachaspati and A. Vilenkin, Phys. Rev. {\bf D30}, 
2036 (1984).

\bibitem{refs} See the papers by D.P. Bennet, F.R. Bouchet,
N.Turok, A. Albrecht, and, E.P.S. Shellard and B. Allen in 
{\it The Formation and Evolution of Cosmic Strings}, eds. 
G.W. Gibbons, S.W. Hawking and T. Vachaspati,
(Cambridge University Press 1990).

\bibitem{sakellariadou} M. Sakellariadou and A. Villenkin,
Phys. Rev. {\bf D42}, 349 (1990)

\bibitem{sawicki} M. J. Sawicki, H. Lin and H. K. L. Yee, Astron. J.,
{\bf 113}, 1 (1997) 

\bibitem{bouchet} See {\it e.g.}~F.R. Bouchet in 
{\it The Formation and Evolution of Cosmic Strings}, eds.
G.W. Gibbons, S.W. Hawking and T. Vachaspati, (Cambridge University
Press 1990).

\bibitem{sakellariadou1} M. Sakellariadou,
Phys. Rev. {\bf D42}, 354 (1990)

\bibitem{hindmarsh} M. Hindmarsh, Phys. Lett. {\bf B251}, 28 (1990)

\end{references}
\end{document}